\documentclass{iau379}

\newcommand{\apj}{ApJ}
\newcommand{\aap}{A{\&}A}
\newcommand{\apjl}{ApJL}
\newcommand{\aj}{AJ}
\newcommand{\jcap}{JCAP}
\newcommand{\mnras}{MNRAS}
\begin{document}

\lefttitle{Y. Jiao et al.}
\righttitle{IAU Symposium 379: Template}

\jnlPage{1}{7}
\jnlDoiYr{2023}
\doival{10.1017/xxxxx}

\aopheadtitle{Proceedings of IAU Symposium 379}
\editors{P. Bonifacio,  M.-R. Cioni, F. Hammer, M. Pawlowski, and S. Taibi, eds.}

\title{Revisiting mass estimates of the Milky Way}

\author{Yongjun Jiao$^1$, François Hammer$^1$, Haifeng Wang$^2$, Jianling Wang$^{1,3}$ and Yanbin Yang$^1$}
\affiliation{$^1$GEPI, Observatoire de Paris, Université PSL, CNRS, Place Jules Janssen, 92195 Meudon, France\\
email:\email{yongjun.jiao@obspm.fr}\\
$^2$CREF, Centro Ricerche Enrico Fermi, Via Panisperna 89A, I-00184 Roma, Italy\\
$^3$CAS Key Laboratory of Optical Astronomy, National Astronomical Observatories, Beijing 100101, China}

\begin{abstract}
We use the rotation curve from \textit{Gaia} data release (DR) 3 to estimate the mass of the Milky Way. We consider an Einasto density profile to model the dark matter component. We extrapolate and obtain a dynamical mass $M=2.75^{+3.11}_{-0.48}\times 10^{11} M_\odot$ at $112$ kpc. This lower-mass Milky Way is consistent with the significant declining rotation curve, and can provide new insights into our Galaxy and halo inhabitants. 
\end{abstract}


\maketitle

\section{Introduction}

The rotation curve (RC) is an essential tool for estimating the enclosed mass at different radii. The Milky Way (MW) disc is in relatively equilibrium because there was no major merger since 9 to 10 Gyr ago \citep{hammer2007,haywood2018}. Then, its dynamical mass can be well established from its RC. Previous RC studies based on the second \textit{Gaia} data release (DR2, \citealp{gaiadr2}) provided a declining MW RC, with a slope of $\beta = -(1.4\pm 0.1)\ \rm km\ s^{-1}\ kpc^{-1}$ and $-(1.7\pm 0.1)\ \rm km\ s^{-1}\ kpc^{-1}$ for \cite{mroz2019} and \cite{eilers2019}, respectively. Such declining RCs have led to estimates of the total MW mass near or well below $10^{12} M_\odot$ \citep{desalas2019,eilers2019,karukes2020,jiao2021}.

The \textit{Gaia} DR3 \citep{gaiadr3} provides a wider sample of stars with new determinations of spectra, radial velocity, chemical abundance, etc. Recently, \citet{wang2023} have applied a statistical inversion method introduced by \cite{lucy1974} to reduce the errors in the distance determination in the Gaia DR3 data-set and provided the MW rotation curve $v_C(R)$ up to about 27.5 kpc. Their RC is in reasonably good agreement with the RC measurement based on \textit{Gaia} DR2 \citep{mroz2019, eilers2019} and DR3 \citep{ou2023}. However, \cite{zhou2023} found a systematically higher RC based on \textit{Gaia} DR3 when it is compared to \cite{wang2023} and \cite{ou2023}. \cite{ou2023} argue that this systematic differences are caused by different methods of measuring distance. In particular, \cite{wang2023} found that a steeper slope of RC, with $\beta = -(2.3\pm 0.2)\ \rm km\ s^{-1}\ kpc^{-1}$, which is smaller than previous studies. A more rapidly declining RC provides a more stringent constraint, leading to a lower mass MW. \cite{jiao2021} have shown that a Navarro–Frenk–White (NFW) profile could not be reconciled with a low mass MW and that an Einasto profile gives a larger mass range. Here we focus on the Einasto profile for the fitting of the MW RC.

\section{Method}

The baryonic model of this study is the same model that we used in \citet[][model I]{jiao2021}, which was obtained from \cite{pouliasis2017} and corresponded to a mass of $\sim9\times 10^{10} M_\odot$. Some studies argue that this model overestimates the mass of baryons towards the outer galactic radii \citep{desalas2019, ou2023}. We have also tested some different baryonic models with smaller mass. Though it has some minor impact on the estimated dark matter (DM) mass, it does not affect the results of dynamical mass (total mass). So the choices of baryonic mass do not change our main results.

We apply the $\chi^2$ method to fit the RC and calculate its associated probability, for which we have tested an extremely large parameter space. The $\chi^2$ is calculated by the sum at each disk radius $R_i$:
\begin{equation}
    \chi^2=\sum_i^N\frac{(v_{\mathrm{mod},i}-v_{\mathrm{obs},i})^2}{\sigma_i^2}
\end{equation}
where $v_{\mathrm{mod}}$ is the modeled circular velocity for the cumulative baryons + DM profiles, $v_{\mathrm{obs}}$ is the observed circular velocity and $\sigma_\mathrm{stat}$ is the statistical uncertainty of the measurement so that $\sigma_{\mathrm{stat},i}=(\sigma^{+}_{v_{\mathrm{obs},i}}+\sigma^{-}_{v_{\mathrm{obs},i}})/{2}$ to which we have added the systematic uncertainty $\sigma_{\mathrm{sys},i}$ to calculate $\sigma_i$. Hence the $\chi^2$ probability can be expressed as :
\begin{equation}
\mathrm{Prob}\left( \frac{ \chi^2}{2}, \frac{N-\nu}{2} \right) = \frac{\gamma\left( \frac{N-\nu}{2}, \frac{ \chi^2}{2} \right)}{\Gamma\left( \frac{N-\nu}{2}\right)}
\end{equation}
where $N$ is the number of independent observed velocity points in the RC and $\nu$ is the number of degrees of freedom.\\
\iffalse
\begin{figure}[h]
    \centering
    \includegraphics[width=\textwidth]{fig/mw_rc_gaia_dr2_with_eina_12.pdf}
    \caption{Rotation curve from \textit{Gaia} DR2 \citep{eilers2019} and Einasto profile fits}
    \label{fit_dr2}
\end{figure}
\fi

\begin{figure}[ht]
    \centering
    \includegraphics[width=\textwidth]{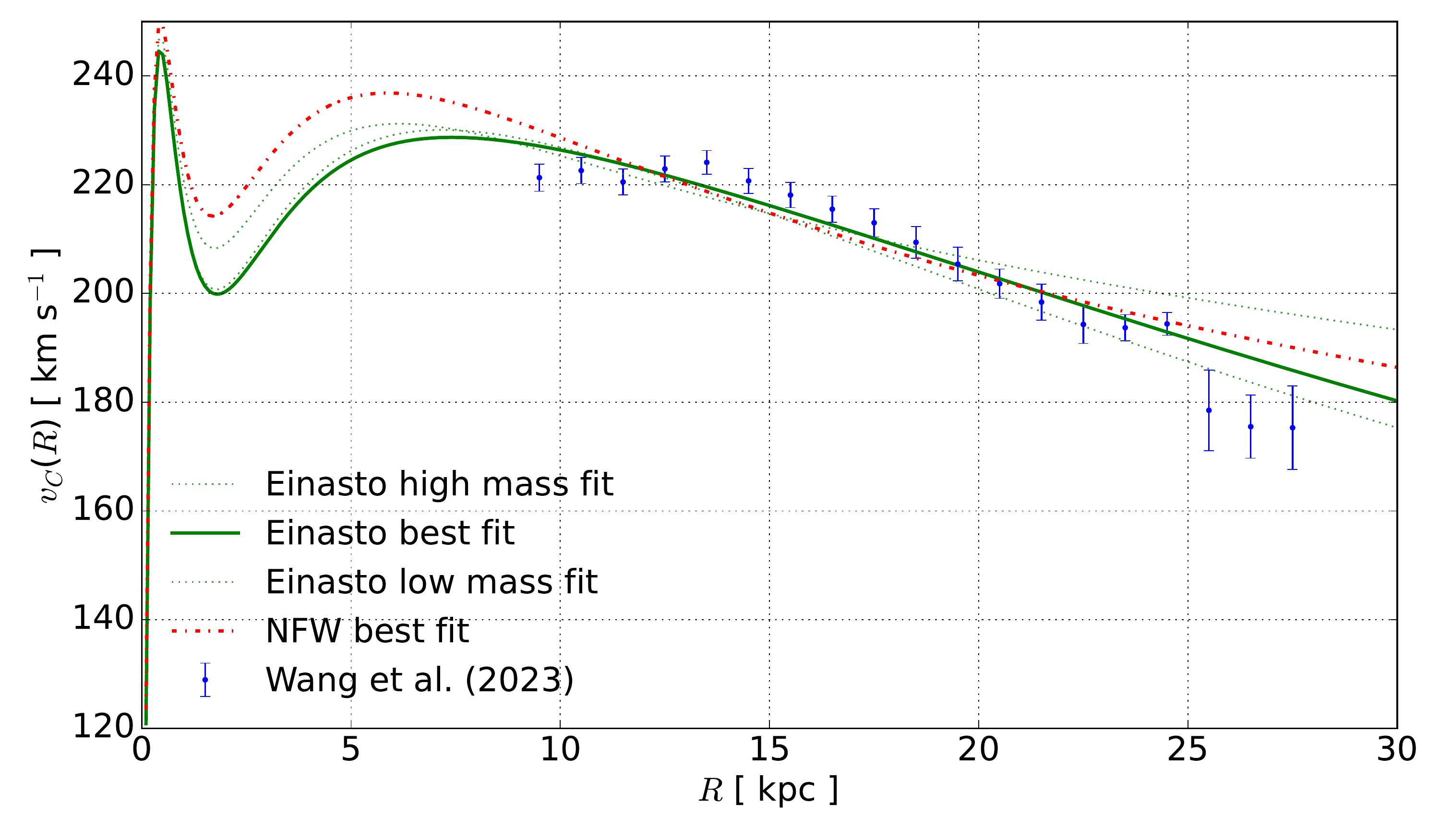}
    \caption{Rotation curve from \textit{Gaia} DR3 \citep[points and error bars][]{wang2023}, Einasto profile fits (solid and dotted lines) and NFW best fit (dash-dotted line)}
    \label{fit_dr3}
\end{figure}

\begin{figure}[ht]
\begin{minipage}[t]{0.5\linewidth}
\centering
    \includegraphics[width=\textwidth]{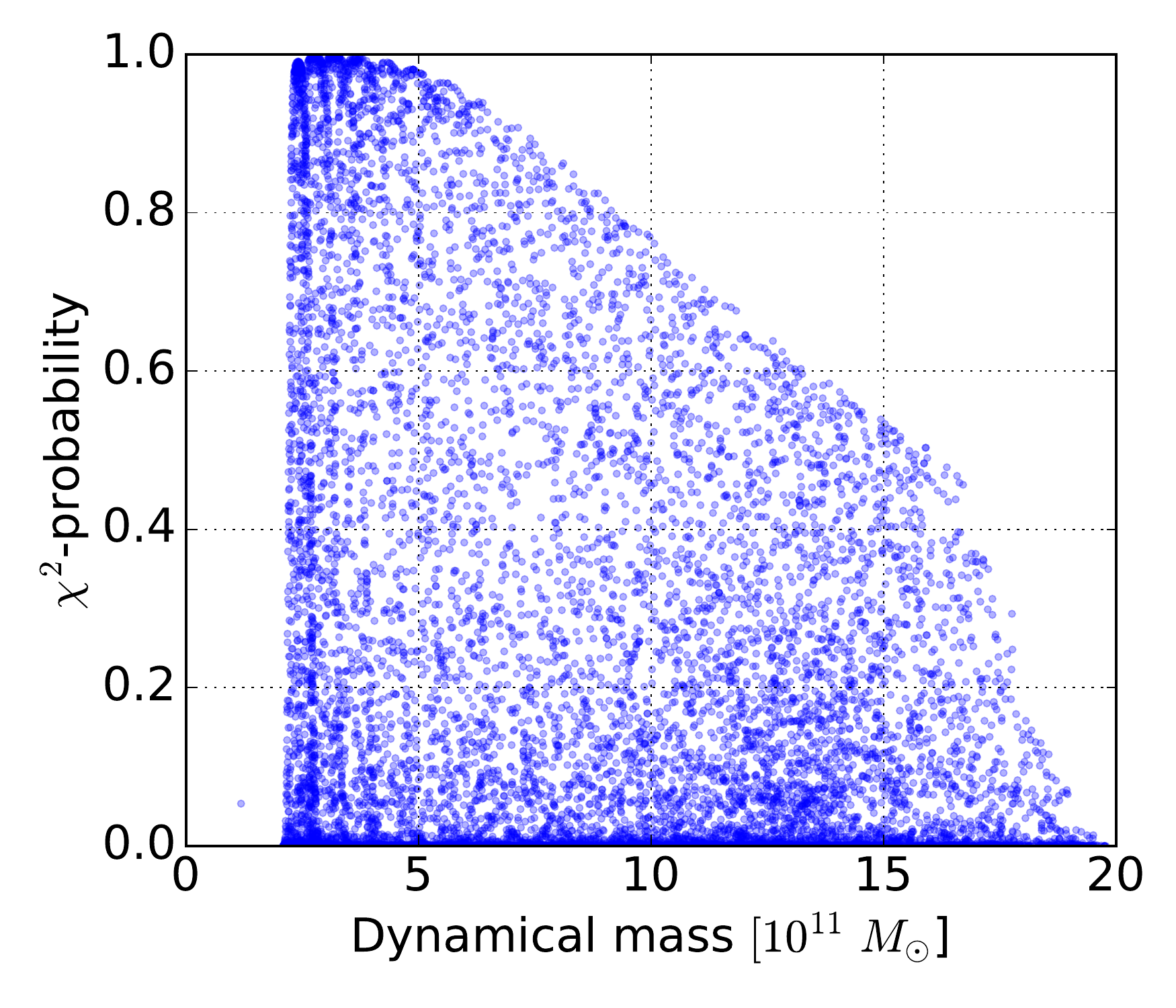}
    %\caption{Estimated dynamical mass and its associated $\chi^2$ probability of Einasto profile fits to RC from \citet{eilers2019}}
    \label{mfit_dr2}
\end{minipage}
\begin{minipage}[t]{0.5\linewidth}
\centering
    \includegraphics[width=\textwidth]{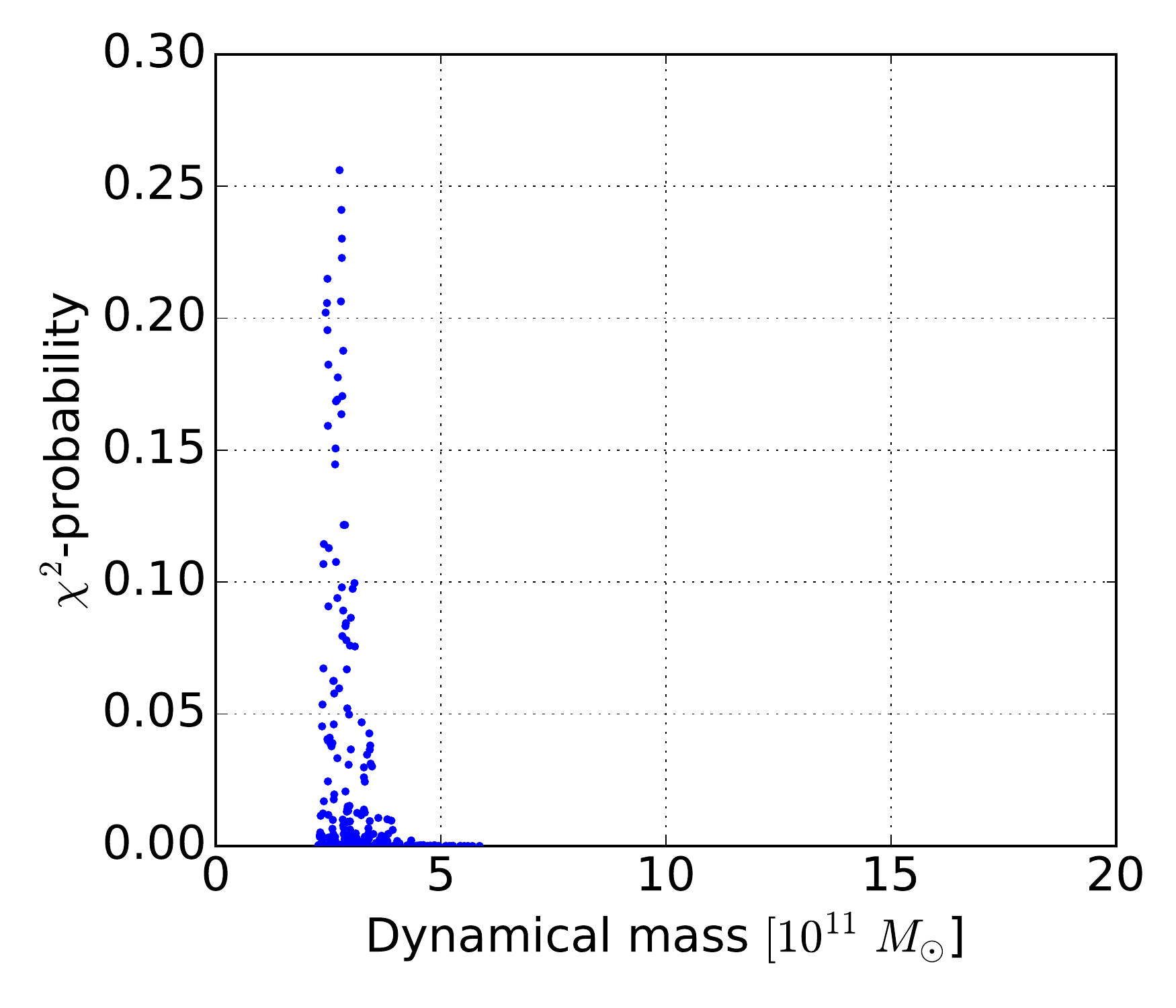}
    %\caption{Estimated dynamical mass and its associated $\chi^2$ probability of Einasto profile fits to RC from \citet{wang2023}}
    \label{mfit_dr3}
\end{minipage}
    \caption{Estimated dynamical mass and its associated $\chi^2$ probability of Einasto profile fits to RC from \cite{eilers2019}(the left panel) and \citet{wang2023}(the right panel).}
    \label{fits}
\end{figure}

\section{Result and conclusion}

\citet{jiao2021} explained in detail that the NFW profile may lead to a possible methodological bias particularly against low MW masses. With the RC from \textit{Gaia} DR3 \citep{wang2023}, we find that the Einasto profile provides better fitting results than the NFW model. In Figure \ref{fit_dr3} we present the RC from \cite{wang2023}, 3 models (low mass fit, best fit and high mass fit) of Einasto profile and best-fit model of NFW profile. The reduced $\chi^2$ of the best fit model is $\sim 1.5$ (for comparison, the reduced $\chi^2$ of best fit NFW profile is $\sim 3.7$). \cite{ou2023} and \cite{labini2023} have also found a similar result. In short, the three-parameter Einasto profile provides a much better fit to the declining RC.

In Fig. \ref{fits}. we compare the results of Einasto profile using \textit{Gaia} DR2 and DR3. We keep the same parameter space as in \cite{jiao2021} but the estimated MW mass range is much narrower. The best-fit dynamical mass remains consistent, from $2.77\times 10^{11}\ M_\odot$ to $2.75\times 10^{11}\ M_\odot$, indicating that the Einasto profile gives a consistent estimated dynamical mass.
%RC of \cite{eilers2019} presents a smaller slope compared to the inner part and 

The low mass fits of RC are mostly constrained by the data at large radii ($>17$ kpc). \cite{wang2023} find a consistent slope beginning at $\sim 13$ kpc. The last five points seem contradictory to each other. For now we think that inconsistencies may be due to the lack of a strict systematic uncertainty analysis, like the neglected cross term in Jeans equation and the uncertainty of scale length (see also \citealt{eilers2019}). We will present this analysis in an upcoming work, but preliminary results show that the systematic does not lead to significant differences in our main results, including the choices of DM profile, estimated mass range and best fit model. 

The Magellanic Clouds (MCs), Globular clusters (GCs) and dwarf galaxies are also widely used as tracers to estimate the enclosed mass of the MW at large radii. However, these methods can easily lead to prior selection bias. \cite{wang2022} found that by excluding two GCs with large orbit energy, Crater and Pyxis, their estimated MW mass decreased from $5.73^{+0.76}_{-0.58}\times 10^{11} M_\odot $ to $5.36^{+0.81}_{-0.68}\times 10^{11} M_\odot$ with Einasto profile. In fact, as relatively old halo inhabitants \citep{hammer2023}, GCs behave consistently with a large MW mass range. We have tested orbits of all GCs with the MW mass from 2 to 15 $\times 10^{11} M_\odot$, and find that almost all GCs are gravitationally bound. However, about half of the dwarf galaxies are not bound with a low mass MW ($\sim 2\times 10^{11} M_\odot$). If most dwarfs came late in the MW halo \citep[$<$ 3 Gyr,][]{hammer2023}, it is unrealistic to use them as tracers for estimating MW mass. For example, it is likely that the consideration of Leo I as a bound satellite would lead to a significant overestimate of the MW mass. On the other hand, orbital studies of other halo inhabitants are sensitive to the choice of MW mass and need more rigorous analysis.

%The mass of MCs is hard to be measured directly. Some studies use the gravitational interactions between the MCs and the stellar disc and halo of the MW to estimate the mass ratio of the MCs to MW. If we assumed a high mass MW model ($\sim 10^{12} M_\odot$), it is reasonably that we have a massive MCs. In this case, the mass of the MCs depends on the MW so it could not be used to constrain the mass of MW anymore. Some analysis of these tracers are based on the MW model and sensitive to the choice of MW mass, they need more rigorous studies.

%
%\bibliographystyle{mnras}
%\bibliography{iau379_paper}

\begin{thebibliography}{}
%
\bibitem[de Salas et al.(2019)]{desalas2019} de Salas, P.~F., Malhan, K., Freese, K., et al.\ 2019, \jcap, 2019, 037. %doi:10.1088/1475-7516/2019/10/037
\bibitem[Eilers et al.(2019)]{eilers2019}Eilers, A.-C., Hogg, D. W., Rix, H.-W., et al.\ 2019, {\apj}, 871,120
\bibitem[Gaia Collaboration et al.(2018)]{gaiadr2} Gaia Collaboration, Brown, A.~G.~A., Vallenari, A., et al.\ 2018, \aap, 616, A1. %doi:10.1051/0004-6361/201833051
\bibitem[Gaia Collaboration et al.(2022)]{gaiadr3} Gaia Collaboration, Vallenari, A., Brown, A.~G.~A., et al.\ 2022, arXiv:2208.00211. %doi:10.48550/arXiv.2208.00211
\bibitem[Hammer et al.(2007)]{hammer2007} Hammer, F., Puech, M., Chemin, L., et al.\ 2007, \apj, 662, 322. %doi:10.1086/516727
\bibitem[Hammer et al.(2023)]{hammer2023} Hammer, F., Li, H., Mamon, G.~A., et al.\ 2023, \mnras, 519, 5059. %doi:10.1093/mnras/stac3758
\bibitem[Haywood et al.(2018)]{haywood2018} Haywood, M., Di Matteo, P., Lehnert, M.~D., et al.\ 2018, \apj, 863, 113. %doi:10.3847/1538-4357/aad235
\bibitem[Jiao et al.(2021)]{jiao2021} Jiao, Y., Hammer, F., Wang, J.~L., et al.\ 2021, \aap, 654, A25. %doi:10.1051/0004-6361/202141058
\bibitem[Karukes et al.(2020)]{karukes2020} Karukes, E.~V., Benito, M., Iocco, F., et al.\ 2020, \jcap, 2020, 033. %doi:10.1088/1475-7516/2020/05/033
\bibitem[Lucy(1974)]{lucy1974} Lucy, L.~B.\ 1974, \aj, 79, 745. %doi:10.1086/111605
\bibitem[Mr{\'o}z et al.(2019)]{mroz2019} Mr{\'o}z, P., Udalski, A., Skowron, D.~M., et al.\ 2019, \apjl, 870, L10. %doi:10.3847/2041-8213/aaf73f
\bibitem[Ou et al.(2023)]{ou2023} Ou, X., Eilers, A.-C., Necib, L., et al.\ 2023, arXiv:2303.12838. %doi:10.48550/arXiv.2303.12838
\bibitem[Pouliasis et al.(2017)]{pouliasis2017} Pouliasis, E., Di Matteo, P., \& Haywood, M.\ 2017, \aap, 598, A66. %doi:10.1051/0004-6361/201527346
\bibitem[Sylos Labini et al.(2023)]{labini2023} Sylos Labini, F., Chrob{\'a}kov{\'a}, {\v{Z}}., Capuzzo-Dolcetta, R., et al.\ 2023, \apj, 945, 3. %doi:10.3847/1538-4357/acb92c
\bibitem[Wang et al.(2022)]{wang2022} Wang, J., Hammer, F., \& Yang, Y.\ 2022, \mnras, 510, 2242. %doi:10.1093/mnras/stab3258
\bibitem[Wang et al.(2023)]{wang2023} Wang, H.-F., Chrob{\'a}kov{\'a}, {\v{Z}}., L{\'o}pez-Corredoira, M., et al.\ 2023, \apj, 942, 12. %doi:10.3847/1538-4357/aca27c
\bibitem[Zhou et al.(2023)]{zhou2023} Zhou, Y., Li, X., Huang, Y., et al.\ 2023, \apj, 946, 73. %doi:10.3847/1538-4357/acadd9
\end{thebibliography}

\begin{discussion}

\discuss{Manuel Bayer}{1. Do I understand correctly that for the modeling of the \citet{eilers2019} data-set you just take into account the uncertainties for the circular velocity?

2. Did you just use the binned data of \citet{eilers2019}?

3. What stars do \citet{wang2023} employ to trace the rotation curve?}

\discuss{Yongjun Jiao}{1. We consider the neglected term in their Jeans equation, which is a cross-term made by the vertical density gradient of the product of the radial and vertical velocities and add an additional systematic uncertainty of $~2\%$ on the velocity scale due to the effect of changing the distance of the Sun to the Galactic center and the scale length, the proper motion of the latter, etc.

2. We directly used their binned data \citep[Table 1]{eilers2019} with overestimated error bars (see above).}

\discuss{Haifeng Wang}{3. \cite{wang2023} use different types of stars in combination with Gaia DR3 line-of-sight velocity, which is based on the Lucy method and mainly in the galactic anticenter direction. The random errors and systematics are acceptable at least for the moment. }

\end{discussion}

\end{document}